\documentclass[
manuscript,
]{revtex4-1}

\usepackage{graphicx}
\usepackage{dcolumn}
\usepackage{bm}
\usepackage{amsmath}

\usepackage[utf8]{inputenc}
\usepackage[labelfont=bf,justification=raggedright]{caption}

\usepackage{xr}

\newcommand{\bra}[1]{\langle #1|}
\newcommand{\ket}[1]{|#1\rangle}
\newcommand{\braket}[2]{\langle #1|#2\rangle}

\newcommand{\ex}[1]{\mathrm{e}^{#1}} 
\newcommand{\abs}[1]{\left|#1\right|}

\newcommand*\df{\mathop{}\!\mathrm{d}}

\renewcommand{\thefigure}{Figure \arabic{figure}}
\renewcommand{\figurename}{}

\begin{document}



\title{Optimal superadiabatic population transfer and gates by dynamical phase corrections}



\author{A. Veps\"al\"ainen}
\address{Low Temperature Laboratory, Department of Applied Physics, Aalto University School of Science, P.O. Box 15100, FI-00076 AALTO, Finland}

\author{S. Danilin}
\address{Low Temperature Laboratory, Department of Applied Physics, Aalto University School of Science, P.O. Box 15100, FI-00076 AALTO, Finland}
\author{G.~S. Paraoanu}
\address{Low Temperature Laboratory, Department of Applied Physics, Aalto University School of Science, P.O. Box 15100, FI-00076 AALTO, Finland}





\begin{abstract}
\bf{In many quantum technologies adiabatic processes are used for coherent quantum state operations, offering inherent robustness to errors in the
control parameters. The main limitation is the long operation time resulting from the requirement of adiabaticity. The superadiabatic method allows for faster operation, by applying counterdiabatic driving that corrects for excitations resulting from the violation of the adiabatic condition. In this article we show how to construct the counterdiabatic Hamiltonian in a system with forbidden transitions by using two-photon processes and how to correct for the resulting time-dependent ac-Stark shifts in order to enable population transfer with unit fidelity. We further demonstrate that superadiabatic stimulated Raman passage can realize a robust unitary NOT-gate between the ground state and the second excited state of a three-level system. The results can be readily applied to a three-level transmon with the ladder energy level structure.}
\end{abstract}

\maketitle


\section{Introduction}
The adiabatic manipulation of a Hamiltonian can be used to control the state of a quantum system in a way which is insensitive to small errors in control parameters \cite{another_reviewSTIRAP,stirap_falci}. The drawback of the adiabatic method is that the changes in the Hamiltonian need to be slow, requiring more time than direct Rabi pulses, thus rendering the system more susceptible to decoherence. This issue is addressed in the superadiabatic or transitionless method by introducing a correction Hamiltonian that cancels the diabatic excitations during the manipulation of the system Hamiltonian \cite{Berry09,Demirplak03}. Thus, it is possible to force the system to follow an adiabatic path even when short transfer times are required. However, the faster the process, the less robust to control parameters the system becomes, eventually leading to a situation where there is little difference between the superadiabatic method and the direct Rabi pulse \cite{sa_stirap_ours}. This trade-off between robustness and speed highlights the role of the superadiabatic method as a bridge between the adiabatic method and direct Rabi pulses. 
Somewhere between these two extrema usually lies the set of parameters that provide optimal performance by achieving balance between decoherence and errors in the controls.


The validity of the superadiabatic method for controlling a two-level system has been demonstrated in several experimental platforms, such as optical lattices \cite{bason12} or trapped ions \cite{Zhang13}. In this article we study the superadiabatic method in the context of a three-level ladder system,
where stimulated Raman adiabatic passage (STIRAP) \cite{stirapfirst,stirap_review_vitanov} can be used to adiabatically excite the system from the ground state $\ket{0}$ to the second excited state $\ket{2}$, enabling robust initial state-preparation. The superadiabatic method requires the realization of a correction Hamiltonian involving complex couplings between all the energy levels of the system. However, in many three-level systems some of the transitions are forbidden, complicating the experimental realization of these couplings. This is the case for example in transmon \cite{transmon_PRA2007}, which is a commonly used qubit type in circuit quantum electrodynamics due to its insensitivity to the charge noise. This system exhibits a ladder energy level structure to which one can apply the STIRAP protocol \cite{stirap_ours}. In a recent experiment that uses nitrogen vacancies in diamonds to create a three-level system \cite{Zhou2017}, the problem of the forbidden transition has been avoided by making the superadiabatic correction in a dressed state basis. In this case the state of the system does not exactly follow the adiabatic path, but the initial and the final states will still be correct. In \cite{Du2016} shortcut to adiabaticity has been applied to STIRAP, which is driven with large single photon detuning, thus reducing the three-level lambda system to an effective two-level system. Here we show that it is also possible to realize a process that exactly follows the adiabatic path using a two-photon process to circumvent the issue of the forbidden transition. The possibility to exploit the two-photon process has been suggested earlier in \cite{arimondo_sa_stirap}, but the ac-Stark shifts introduced by the two-photon driving have not been considered so far. Also, in the context of STIRAP, protocols involving two-photon physics have been proposed \cite{stefano_stirap}. The additional pulse acting on the 0\---2 transition can be avoided by using invariant-based engineering \cite{chen_muga_2012}, but there the maximum achievable fidelity is limited by the boundary condition, which ensures that the pulse amplitudes remain finite. The method proposed here does not have such limitation. On the other hand, it is also possible to apply the rapid adiabatic passage protocol \cite{malinovsky2001general,Arkhipkin2003} on the two-photon transition, which eliminates the need of pulses on 0\---1 and 1\---2 transitions, but in this case the weak two-photon coupling increases the duration of the protocol. In superadiabatic STIRAP the counterdiabatic correction pulse applied on the two-photon transition requires significantly smaller amplitude with area corresponding to a $\pi$ pulse, and therefore does not impose severe limitations on the speed of the protocol.

Here we show how to modify the superadiabatic process to take ac-Stark shifts into account and demonstrate that under these conditions superadiabatic STIRAP (saSTIRAP) can be used to realize a unit fidelity population transfer between the ground state and the second excited state of the system.  A related method based on derivative removal by adiabatic gate \cite{drag_wilhelm} has been used to eliminate leakage of the drives in weakly anharmonic systems. 
We further employ our saSTIRAP protocol based on two-photon counterdiabatic driving to demonstrate a robust NOT gate in a three-level ladder system. Typically, the Hamiltonians used for adiabatic transfer depend on the initial state $\ket{\psi_{\rm i}} = \ket{a}$ and the target state $\ket{\psi_{\rm f}} = \ket{b}$. To reverse the process, as needed for a NOT operation, a different Hamiltonian would be required, preventing one from constructing a unitary that acts on arbitrary states. However, we show here that it is possible to find a Hamiltonian which transfers an unknown state
$\ket{\psi_i} = \alpha\ket{a} + \beta\ket{b}$ to $\ket{\psi_{\rm f}} = \beta\ket{a} + \alpha\ket{b}\ex{-i\phi_{\rm f}}$ with $|\alpha|^2 + |\beta|^2 = 1$, thus realizing a NOT gate apart from a phase factor $\phi_{\rm f}$. STIRAP with single photon detuning is known to have this property \cite{du2014experimental}, and in this article we show how to extend the concept to saSTIRAP. Furthermore, if the STIRAP part is split into two fractional STIRAPs (f-STIRAP) \cite{fractional_stirap} with the reversed pulse order, it is possible to cancel the phase factor $\phi_{\rm f}$, enabling the process to be used as a robust NOT gate \cite{fractional_gate_stirap}. Superadiabatic gate would provide an important advantage over STIRAP, because its fidelity does not suffer from the applied detuning. We also show that the superadiabatic NOT gate is robust with respect to its control parameters by comparing its performance to a diabatic $\pi$-pulse.

\section{Results}



\begin{figure}[tb]
	\centering
	\includegraphics[width=1.0\textwidth]{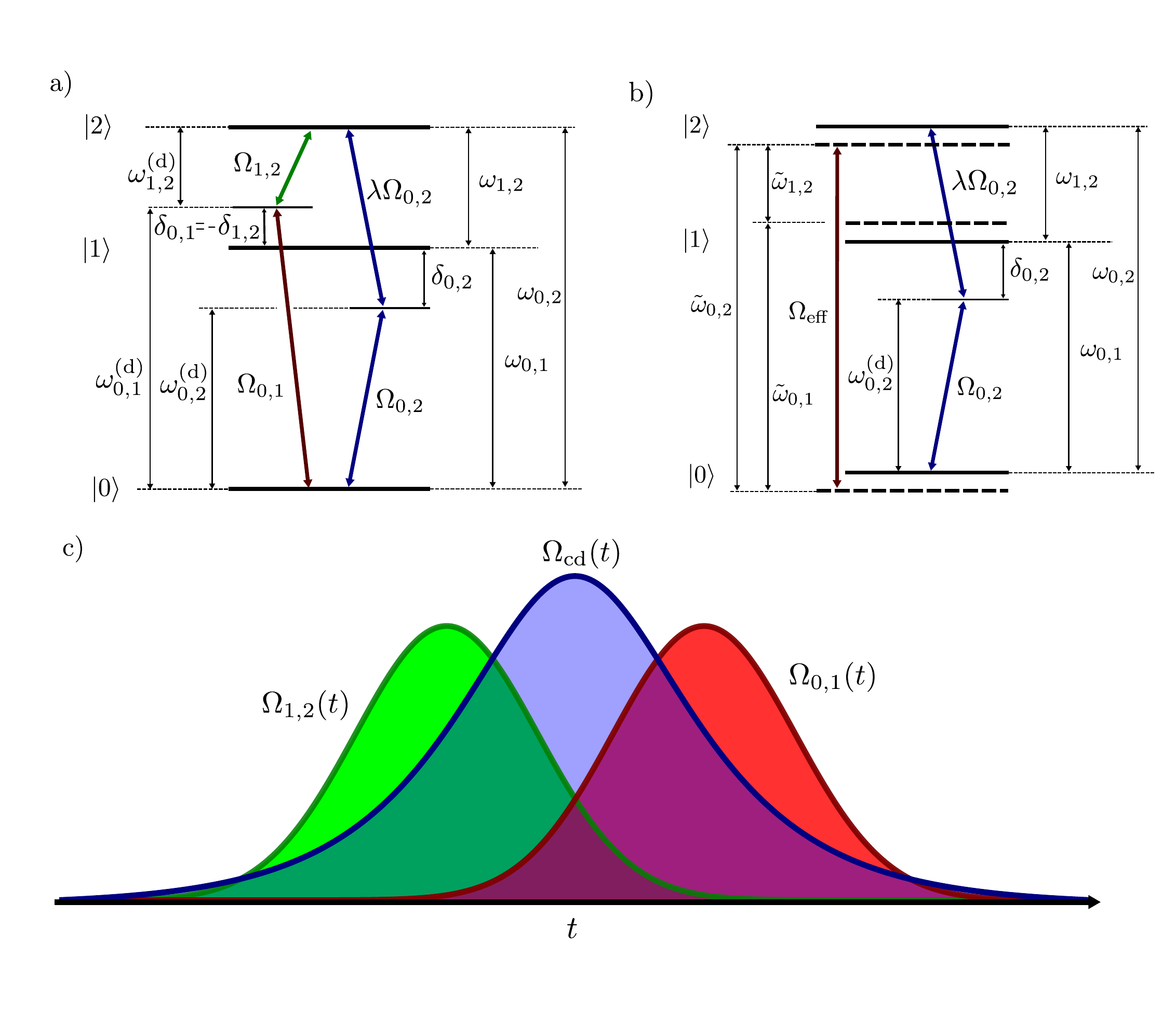}
	\caption{a) The energy level diagram of the three-level system with transition energies $\hbar\omega_{0,1}$, $\hbar\omega_{1,2}$ and $\hbar\omega_{0,2}$. The system is driven by three tones with frequencies $\omega_{0,1}^{\rm (d)}$, $\omega_{1,2}^{\rm (d)}$ and $\omega_{0,2}^{\rm (d)}$, which are detuned from their corresponding transitions by $\delta_{0,1}$, $\delta_{1,2}$ and $\delta_{0,2}$. In this article we have assumed that the two-photon resonance condition is satisfied and therefore $\delta_{0,1} = -\delta_{1,2}$. The amplitudes of the drives are given by $\Omega_{0,1}$, $\Omega_{1,2}$ and $\Omega_{0,2}$. b) The two-photon driving creates ac-Stark shifts to the energy levels. The transition energies of the shifted levels are given by $\hbar\tilde{\omega}_{0,1}$, $\hbar\tilde{\omega}_{1,2}$, and $\hbar\tilde{\omega}_{0,2}$ and the effective 0\---2 coupling is given by $\Omega_{\rm eff}$. c) The pulse shapes used in saSTIRAP.}
	\label{fig:diagram}
\end{figure}

In a three-level ladder system STIRAP can be used to transfer population from the ground state $\ket{0}$ to the second excited state $\ket{2}$ without populating the intermediate state $\ket{1}$ during the process. This is achieved with two pulses driving the 0\---1 transition and 1\---2 transition, respectively. In order to satisfy the adiabatic condition, the amplitudes of the pulses have to change slowly with respect to each other and they need to be applied in the counterintuitive order, {\it i.e}, the 1 \--- 2 pulse comes before 0 \--- 1 pulse, with a small overlap. A common choice for the pulse shapes is to use Gaussian envelopes,

\begin{equation}
\label{eq:pulse_shapes}
\begin{aligned}
\Omega_{0,1}(t) &= \Omega_{0,1}\exp{\left[\frac{-t^2}{2\sigma^2}\right]}, \\
\Omega_{1,2}(t) &= \Omega_{1,2}\exp{\left[\frac{-(t-t_{\rm s})^2}{2\sigma^2}\right]},
\end{aligned}
\end{equation}
with real amplitudes $\Omega_{i,j}$ and overlap $t_{\rm s}$. With this parametrization the negative values of the overlap correspond to the counterintuitive case. The three-level system with the two drives can be described by the Hamiltonian

\begin{equation}
\label{eq:stirap_ham}
H_f = \hbar\sum_{j=0}^2 \omega_j + \frac{\hbar}{2}\sum_{j=0}^1\Omega_{j,j+1}(t)\sigma_{j,j+1}^+\exp{(-i\omega_{j,j+1}^{(d)}t)} + {\rm h.c},
\end{equation}
which after moving to a doubly rotating frame at the drive frequencies becomes

\begin{equation}
H = \frac{\hbar}{2}\begin{bmatrix}
0 & \Omega_{0,1}(t) & 0 \\
\Omega_{0,1}(t) & 2\delta_{0,1} & \Omega_{1,2}(t) \\
0 & \Omega_{1,2}(t) & 2(\delta_{0,1}+\delta_{1,2})
\end{bmatrix},
\end{equation}
where $\delta_{0,1} = \omega_{0,1}^{\rm (d)} - \omega_{0,1}$ and $\delta_{1,2} = \omega_{1,2}^{\rm (d)} - \omega_{1,2}$ are the detunings of the drives from their corresponding transitions and the rotating wave approximation has been used to neglect the fast rotating terms. A diagram of the drives is presented in \ref{fig:diagram}a). The two-photon resonance assumption $\delta_{0,1} + \delta_{1,2} = 0$ allows one to diagonalize the Hamiltonian giving a simple form for the instantaneous eigenstates of the system
\begin{equation}
\begin{aligned}
\ket{+} =& \sin\Phi\ket{B} + \cos\Phi\ket{1}, \\ 
\ket{-} =& \cos\Phi\ket{B} - \sin\Phi\ket{1}, \\
\ket{D} =& \cos\Theta\ket{0} - \sin\Theta\ket{2},
\end{aligned}
\label{eq:stirap_eig}
\end{equation}
where $\ket{B} = \sin\Theta\ket{0} + \cos\Theta\ket{2}$ and the eigenvalues of the Hamiltonian are $\hbar\omega_+ = \hbar\left(\delta_{0,1} + \sqrt{\delta_{0,1}^2 + \Omega_{0,1}(t)^2 + \Omega_{1,2}(t)^2}\right)$, $\hbar\omega_- = \hbar\left(\delta_{0,1} - \sqrt{\delta_{0,1}^2 + \Omega_{0,1}(t)^2 + \Omega_{1,2}(t)^2}\right)$, and $\hbar\omega_D = 0$. The mixing angle $\Theta$ is defined as $\tan\Theta(t) = \Omega_{0,1}(t)/\Omega_{1,2}(t)$, while 
\begin{equation}
\tan\Phi(t) = \frac{\sqrt{\Omega_{0,1}(t)^2 + \Omega_{1,2}(t)^2}}{\sqrt{\Omega_{0,1}(t)^2 + \Omega_{1,2}(t)^2 + \delta_{0,1}^2} + \delta_{0,1}}.
\end{equation}
This simplifies to $\Phi(t) = \pi/4$ in the case of resonant driving, $\delta_{0,1} = \delta_{1,2} = 0$. Initially, the system starts from the ground state $\ket{0}$, which corresponds to the instantaneous state $\ket{D}$ because $\lim_{t \to -\infty} \Theta(t) = 0$. As $\Omega_{0,1}(t)$ and $\Omega_{1,2}(t)$ evolve in time, the mixing angle $\Theta$ reaches $\pi/2$, yielding $\ket{D(\infty)} = -\ket{2}$. As long as the changes in the Hamiltonian are slow enough to fulfill the adiabatic condition 
\begin{equation}
|\bra{\pm}(\mathrm{d/d}t)\ket{D}| \ll |\omega_\pm-\omega_D|
\label{eq:adiabatic_condition}
\end{equation}
there are no diabatic excitations from the state $\ket{D}$ to either of the states $\ket{-}$ or $\ket{+}$, thus realizing the required population inversion. The condition can be simplified using the global adiabatic condition, defined as the time integral of Eq. \eqref{eq:adiabatic_condition}. For the Gaussian pulses with $\Omega_{0,1} \approx \Omega_{1,2} \approx \Omega$, the integration gives $\sigma\Omega \ll 1$ \cite{stirap_ours}. Usually the maximum available pulse amplitude is limited by the experimental conditions or the lack of anharmonicity in the energy levels of the system, which leaves the duration and the separation of the Gaussian pulses as the only remaining control parameters. According to the adiabatic condition, longer pulses lead to smaller diabatic excitations resulting in higher fidelity for the operation. In practice, the optimal pulse length is limited by the decoherence affecting the system.


The superadiabatic method allows for compensation of the process's non-adiabaticity by introducing additional drive pulses which exactly cancel the unwanted diabatic excitations. In this way the system state always follows its instantaneous state in the diagonalized basis, even if the adiabatic condition is not fully satisfied. With this method it is in principle possible to make the adiabatic evolution of the system arbitrarily fast without losing in fidelity, but as it has been previously shown, this will come at the cost of robustness in the control parameters \cite{sa_stirap_ours}. 
The form for the counterdiabatic Hamiltonian that reverses the non-adiabatic evolution of the system is given by (see e.g. \cite{arimondo_sa_stirap})

\begin{equation}
\label{eq:re_ham}
H_{\rm cd}(t) = i\hbar \sum_n\ket{\partial_t n(t)}\bra{n(t)} - \braket{n(t)}{\partial_t n(t)}\ket{n(t)}\bra{n(t)},
\end{equation}
where $\ket{n}$ are the instantaneous eigenstates of the system from Eq. \eqref{eq:stirap_eig}. For Gaussian STIRAP pulses with equal amplitudes this leads to

\begin{equation}
H_{\rm cd}(t) = i\hbar
\begin{bmatrix}
0&0&\dot{\Theta}(t) \\
0&0&0 \\
-\dot{\Theta}(t)&0&0
\end{bmatrix},
\label{eq:cd_ham}
\end{equation}
with
\begin{equation}
2i\dot{\Theta}(t) = \Omega_{\rm cd}(t) = 2\frac{\dot{\Omega}_{0,1}(t)\Omega_{1,2}(t) - \Omega_{0,1}(t)\dot{\Omega}_{1,2}}{\Omega_{0,1}(t)^2 + \Omega_{1,2}(t)^2}\exp{(i\pi/2)} = -\frac{t_{\rm s}}{\sigma^2}\frac{\exp{(i\pi/2)}}{\cosh\left[-\frac{t_{\rm s}}{\sigma^2}\left(t - t_{\rm s}/2\right)\right]}.
\label{eq:cd}
\end{equation}
The pulse shapes used in saSTIRAP are shown in \ref{fig:diagram}c). The counterdiabatic Hamiltonian implies that realizing the counterdiabatic correction requires coupling between the states $\ket{0}$ and $\ket{2}$. In many experimental configurations, for example in superconducting transmon, the direct 0 \--- 2 transition is forbidden by the symmetry of the eigenstates \cite{transmon_PRA2007}. However, it is possible to use a two-photon process to create an effective 0 \--- 2 coupling by driving the system at $\omega_{0,2}^{(\rm d)} = \omega_{0,2}/2$, see \ref{fig:diagram}a). In addition to creating the 0 \--- 2 coupling, the two-photon drive also introduces ac-Stark shifts on the energy levels, which modifies the effective Hamiltonian as depicted in \ref{fig:diagram}b). This implies that the Hamiltonian \eqref{eq:cd_ham} is not exactly reproduced by the two-photon scheme.
Next, we study the effect of the two-photon driving on saSTIRAP and develop a method for correcting the counter-diabatic Hamiltonian \eqref{eq:cd_ham} accordingly, enabling unit fidelity population transfer.

\subsection{Dynamical phase correction}

We get the Hamiltonian governing the saSTIRAP with the two photon drive from Eq. \eqref{eq:stirap_ham} by adding an additional drive term

\begin{equation}
\label{eq:two_photon_ham}
H_{\rm 2ph}^{\rm d} = \frac{\hbar}{2} \left(\sigma^+_{0,1}\Omega_{0,2}^*(t)\ex{-i\omega_{0,2}^{(d)}t} + \lambda \sigma^+_{1,2}\Omega_{0,2}^*(t) \ex{-i\omega_{0,2}^{(d)}t}\right) + {\rm H.c.},
\end{equation}
which couples simultaneously to both 0 \--- 1 and 1 \--- 2 transitions. The factor $\lambda$ takes into account the fact that the drive couples stronger to the higher transitions. For a transmon $\lambda\approx\sqrt{2}$, which follows from its almost harmonic energy level structure.
Next we write the Hamiltonian for the three-level system and the two-photon drive in a frame rotating with $\omega_{0,2}^{(d)}$ in order to calculate the effective two-photon coupling. The Hamiltonian reads

\begin{equation}
\label{eq:two_photon_qubit}
H_{\rm 2ph} = 
\frac{\hbar}{2}\begin{bmatrix}
0&\Omega_{0,2}(t)&0 \\
\Omega_{0,2}^*(t)&2\delta_{0,2}&\lambda\Omega_{0,2}(t) \\
0&\lambda\Omega_{0,2}^*(t)& 4\delta_{0,2} - 2\Delta
\end{bmatrix},
\end{equation}
where $\delta_{0,2} = \omega_{0,1} - \omega_{0,2}^{(d)}$ and $\Delta = \omega_{0,1} - \omega_{1,2}$ is the anharmonicity of the system. We can solve the effective 0\---2 transition rate $\Omega_{\rm eff}$ using adiabatic elimination \cite{adiabatic_elimination} by writing the Schrödinger equation for the Hamiltonian in Eq. \eqref{eq:two_photon_qubit} as

\begin{equation}
\begin{aligned}
i\dot{\alpha} =& \frac{\beta\Omega_{0,2}}{2}, \\
i\dot{\beta} =& \frac{\alpha\Omega_{0,2}^*}{2} + \beta\delta_{0,2} + \frac{\gamma \lambda\Omega_{0,2}}{2}, \\
i\dot{\gamma} =& \frac{\beta \lambda\Omega_{0,2}^*}{2} + \gamma(2\delta_{0,2}-\Delta),
\end{aligned}
\label{eq:schrodinger}
\end{equation}
where the state of the system is $\ket{\psi(t)} = \alpha\ket{0} + \beta\ket{1} + \gamma\ket{2}$. By assuming $\delta_{0,2} \gg |\Omega_{0,2}|$ the population of state $\ket{1}$ is almost constant, allowing us to write $\dot{\beta} = 0$, resulting in

\begin{equation}
\beta = -\frac{\alpha\Omega_{0,2}^* + \gamma \lambda \Omega_{0,2}}{2\delta_{0,2}}.
\end{equation}
Substitution back to Eq. \eqref{eq:schrodinger} leads to an effective Hamiltonian

\begin{equation}
H_{\rm ae} = 
\frac{\hbar}{2}\begin{bmatrix}
-\frac{|\Omega_{0,2}|^2}{2\delta_{0,2}}&0&-\frac{\lambda\Omega_{0,2}^2}{2\delta_{0,2}} \\
0&0&0 \\
-\frac{\lambda\Omega_{0,2}^{*2}}{2\delta_{0,2}}&0&4\delta_{0,2}-2\Delta -\frac{\lambda^2|\Omega_{0,2}|^2}{2\delta_{0,2}}
\end{bmatrix},
\end{equation}
with the coupling 
\begin{equation}
\Omega_{\rm eff} = -\frac{\lambda\Omega_{0,2}^2}{2\delta_{0,2}}
\label{eq:effective_coupling}
\end{equation}
for the 0-2 transition. This Hamiltonian only acts in the subspace $\{\ket{0},\ket{2}\}$, thus validating the use of a two-photon process for creating the 0-2 coupling.
However, the method of adiabatic elimination does not take into account the ac-Stark shift which is induced to state $\ket{1}$ due to the two-photon drive. To calculate this contribution, we employ perturbation theory to estimate the shifts on all the 3 states.
We take the diagonal of the Hamiltonian in Eq. \eqref{eq:two_photon_qubit} as $H_0$ with eigenenergies $E_n$ and consider the off-diagonal elements $V = \frac{\hbar\Omega_{0,2}(t)}{2}[\ket{0}\bra{1} + \lambda\ket{1}\bra{2}] + {\rm h.c}$ as a small perturbation, where $\Omega_{0,2}(t)$ is assumed to change slowly in time, justifying the use of time-independent perturbation theory. Up to the second order the perturbative correction to the energy levels is given by

\begin{equation}
\label{eq:perturbation}
\tilde{E}_n = E_n + \bra{n}V\ket{n} + \sum_{k \neq n} \frac{|\bra{k}V\ket{n}|^2}{E_n - E_k},
\end{equation}
and the resulting shifts in the energies $\hbar\epsilon_n = \tilde{E}_n - E_n$ are 

\begin{equation}
\begin{aligned}
\epsilon_0(t) &= -\frac{|\Omega_{0,2}(t)|^2}{4\delta_{0,2}}, \\
\epsilon_1(t) &= \frac{|\Omega_{0,2}(t)|^2}{4\delta_{0,2}} + \frac{\lambda^2|\Omega_{0,2}(t)|^2}{4(\Delta - \delta_{0,2})}, \\
\epsilon_2(t) &= - \frac{\lambda^2|\Omega_{0,2}(t)|^2}{4(\Delta - \delta_{0,2})},
\end{aligned}
\end{equation}
leading to shifts in the transition frequencies
\begin{equation}
\label{eq:energy_shifts}
\begin{aligned}
\epsilon_{0,1}(t) &= \frac{|\Omega_{0,2}(t)|^2}{2\delta_{0,2}} + \frac{\lambda^2|\Omega_{0,2}(t)|^2}{4(\Delta - \delta_{0,2})}, \\
\epsilon_{1,2}(t) &=  - \frac{|\Omega_{0,2}(t)|^2}{4\delta_{0,2}} - \frac{\lambda^2|\Omega_{0,2}(t)|^2}{2(\Delta - \delta_{0,2})}, \\
\epsilon_{0,2}(t) &= \frac{|\Omega_{0,2}(t)|^2}{4\delta_{0,2}} - \frac{\lambda^2|\Omega_{0,2}(t)|^2}{4(\Delta - \delta_{0,2})}.
\end{aligned}
\end{equation}
The shifts create a time varying detuning for all the three drives which impairs the performance of the saSTIRAP protocol. In the case $\Omega_{0,2}(t)$ is constant in time it is possible to simply correct for the error by introducing detunings to the drives corresponding to the shifts given in Eqs. \eqref{eq:energy_shifts}. For the saSTIRAP protocol this is not possible because the pulse envelopes are time varying functions. We can, however, introduce a dynamical phase correction to the pulses, which evolves with the two-photon drive amplitude. Somewhat related strategies of chirping the pulses have been used also in STIRAP \cite{stefano_stirap}. We start by writing the effective Hamiltonian for all the 3 pulses in a doubly rotating frame rotating with $\omega_{0,1}^{\rm (d)}$ and $\omega_{1,2}^{\rm (d)}$. The Hamiltonian is given by

\begin{equation}
H_{\rm eff} = 
\frac{\hbar}{2}\begin{bmatrix}
2\epsilon_0(t)&\Omega_{0,1}(t){\mathrm e}^{i\tilde{\phi}_{0,1}(t)}&\Omega_{\rm eff}(t){\rm e}^{i\tilde{\phi}_{0,2}(t)} \\
\Omega_{0,1}(t){\mathrm e}^{-i\tilde{\phi}_{0,1}(t)}&2[\epsilon_1(t) + \delta_{0,1}]&\Omega_{1,2}{\mathrm e}^{i\tilde{\phi}_{1,2}(t)} \\
\Omega_{\rm eff}^*(t){\rm e}^{-i\tilde{\phi}_{0,2}(t)}&\Omega_{1,2}(t){\mathrm e}^{-i\tilde{\phi}_{1,2}(t)}&2[\epsilon_{2}(t) + \delta_{0,1} + \delta_{1,2}]
\end{bmatrix},
\label{eq:H_eff}
\end{equation}
where $\tilde{\phi}_{0,1}(t)$, $\tilde{\phi}_{1,2}(t)$ and $\tilde{\phi}_{0,2}(t)$ are the dynamical correction phases defined as

\begin{equation}
\begin{aligned}
\tilde{\phi}_{0,1}(t) &= \int_{-\infty}^t\! \epsilon_{0,1}(\tau)\,{\rm d}\tau, \\
\tilde{\phi}_{1,2}(t) &= \int_{-\infty}^t\! \epsilon_{1,2}(\tau)\,{\rm d}\tau, \\
\tilde{\phi}_{0,2}(t) &= \int_{-\infty}^t\! \epsilon_{0,2}(\tau)\,{\rm d}\tau,
\end{aligned}
\label{eq:phi_correction}
\end{equation}
so that the detunings caused by the two-photon pulse are cancelled. 
Next, we compare the approximations calculated in Eqs. \eqref{eq:energy_shifts} to a numerical simulation where we can probe the ac-Stark shifted transition frequencies of the qutrit under a constant two-photon drive. The simulation is performed by numerically finding the time evolution of the system from

\begin{figure}[tb]
	\centering
	\includegraphics[width=1.0\textwidth]{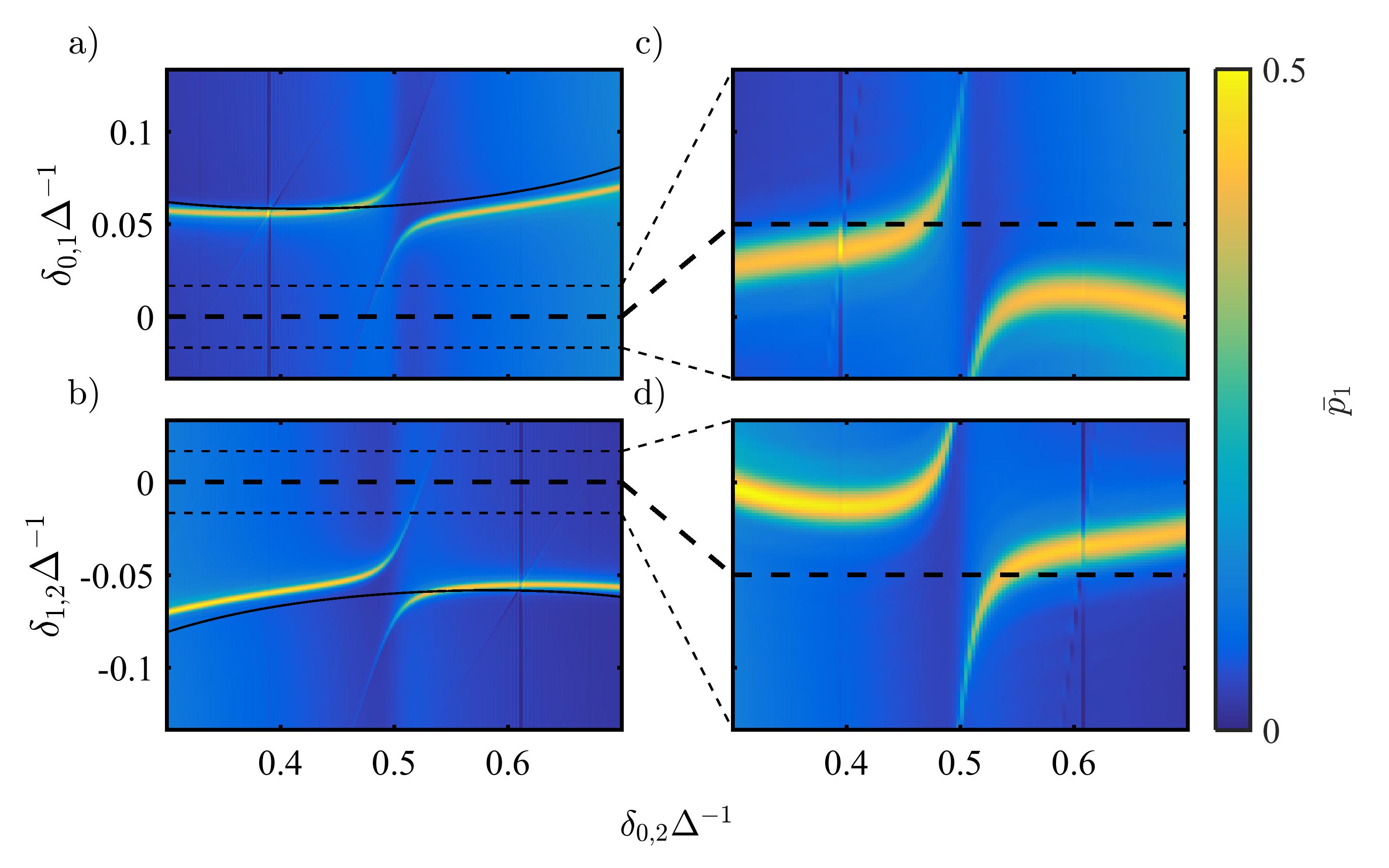}
	\caption{a) Averaged population of state $\ket{1}$, $\bar{p}_1$, as a function of the two-photon drive detuning $\delta_{0,2}$ and the 0 \--- 1 probe detuning $\delta_{0,1}$. When the probe frequency $\omega_{0,1}^{\rm (d)} =  \omega_{0,1} + \delta_{0,1}$ is resonant with the ac-Stark shifted 0\---1 transition frequency $\tilde{\omega}_{0,1}$, a peak is obtained in the state $\ket{1}$ population. The solid black line is the perturbation theory estimate of the ac-Stark shift, $\epsilon_{0,1}$, given in Eq. \eqref{eq:energy_shifts}. b) A similar simulation as in a), but the ac-Stark shift of the 1\---2 transition is probed instead. Initially the system is prepared in state $\ket{2}$ so that when the 1\---2 probe frequency $\omega_{1,2}^{\rm (d)} =  \omega_{1,2} + \delta_{1,2}$ is resonant with $\tilde{\omega}_{1,2}$, state $\ket{1}$ gets populated. Panels c) and d) show the corresponding simulations with the dynamical phase corrections given in Eq. \eqref{eq:phi_correction} applied to the Hamiltonian \eqref{eq:H_eff}, which results in the effective cancellation of the ac-Stark shifts. In all the figures the avoided crossover structure is due to the two-photon drive becoming resonant with the 0\---2 transition. In the simulation we have taken $\lambda = 1$ and $\abs{\Omega_{0,2}} = 0.2\Delta$.
	}
	\label{fig:dw02_dw}
\end{figure}

\begin{figure}[tb]
	\centering
	\includegraphics[width=1.0\textwidth]{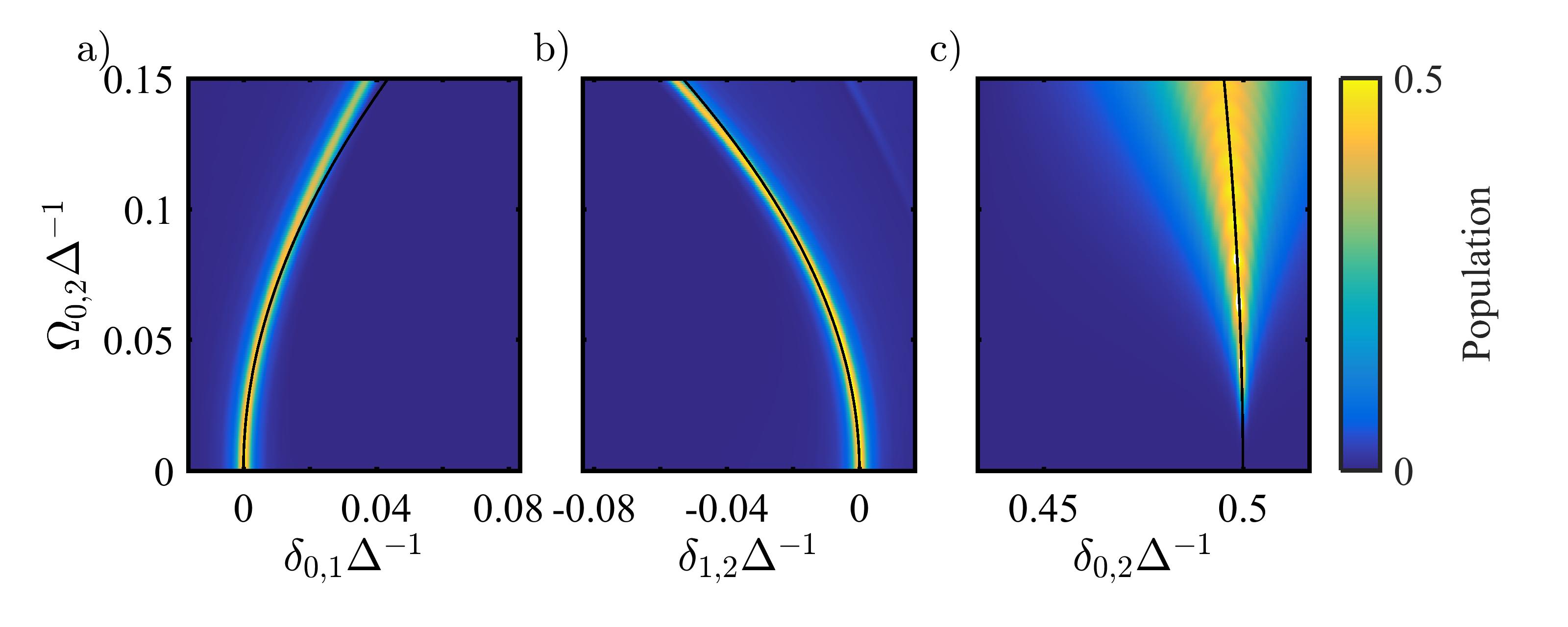}
	\caption{a) Averaged population $\bar{p}_1$ as a function of the 0 \--- 1 probe detuning $\delta_{0,1}$ and the two-photon drive amplitude $\Omega_{0,2}$. Initially the system is in the ground state $\ket{0}$. The solid line is the ac-Stark shift $\epsilon_{0,1}$ given in Eq. \eqref{eq:energy_shifts}. b) Averaged population $\bar{p}_1$ as a function of the 1 \--- 2 probe detuning $\delta_{1,2}$. The initial state of the system is $\ket{2}$. The solid line shows $\epsilon_{1,2}$. c) Averaged population $\bar{p}_2$ when the amplitude and the detuning of the two-photon drive are varied. The solid line shows half of the ac-Stark shift $\epsilon_{0,2}/2$. The simulation is performed with the factor $\lambda = \sqrt{2}$.
	}
	\label{fig:dw_A}
\end{figure}

\begin{equation}
\label{eq:master_eq}
i\hbar\ket{\dot{\psi}} = H_{\rm sim}\ket{\psi}
\end{equation}
with the Hamiltonian $H_{\rm sim} = H_{\rm f} + H_{\rm 2 ph}^{\rm d}$, where $H_{\rm f}$ and $H_{\rm 2phi}^{\rm d}$ are given in Eqs. (\ref{eq:stirap_ham}, \ref{eq:two_photon_ham}). For solving the differential equation we employ the Runge-Kutta algorithm with adaptive step size. The state of the system is given by $\ket{\psi} = \alpha\ket{0} + \beta\ket{1}+\gamma\ket{2}$ along with the occupation probabilities of the states $p_0 = |\alpha|^2$, $p_1 = |\beta|^2$, and $p_2 = |\gamma|^2$. In the simulation we fix the two-photon drive amplitude at $|\Omega_{0,2}|/\Delta$ = 0.2, take $\lambda = 1$, and we employ a second drive to probe the shifts in the transition frequencies $\epsilon_{0,1} = \tilde{\omega}_{0,1} - \omega_{0,1}$ and $\epsilon_{1,2} = \tilde{\omega}_{1,2} - \omega_{1,2}$. For the rest of the article we use the anharmonicity of the system $\Delta$ as the scaling factor for all the parameters because that determines the maximum amplitudes for all the pulses and thus sets the relevant time scale of the system. In \ref{fig:dw02_dw}a) we present the results of the simulation by showing the population of state $\ket{1}$ as a function of the probe frequency, described by the detuning $\delta_{0,1}$ from the bare 0 \--- 1 transition frequency $\omega_{0,1}$. The state $\ket{1}$ gets populated only when the probe is resonant with the ac-Stark shifted frequency $\tilde{\omega}_{0,1}$, allowing us to numerically extract $\epsilon_{0,1}$ as the shift of the peak from $\delta_{0,1} = 0$. According to Eq. \eqref{eq:energy_shifts}, $\epsilon_{0,1}$ also depends on the two-photon detuning $\delta_{0,2}$ which we show on the horizontal axis of \ref{fig:dw02_dw}a). The solid line in the figure corresponds to the perturbation theory approximation given in Eq. \eqref{eq:energy_shifts}, which is in a good agreement with the numerical result, justifying the use of the approximation to describe the ac-Stark shifts. In case $\delta_{0,2} = \Delta/2$, the two-photon drive is in resonance with the 0 \--- 2 transition, resulting in an avoided crossing in the spectrum due to the formation of dressed states, similarly as in the Autler-Townes effect \cite{autler_townes_sillanpaa}. This is not related to the ac-Stark shift caused by the off-resonant driving of 0 \--- 1 and 1 \--- 2 transitions, and would also be present in the spectrum under a direct 0 \--- 2 coupling in the Hamiltonian. This feature is not captured by the perturbation theory approximations $\epsilon_{i,j}$.

The probe amplitude used in the simulation is chosen to be $\Omega_{0,1} =0.001\,\Delta \ll \Omega_{0,2}$ so that the probe itself does not perturb the system. Respectively, the duration of the simulation is $t_{\rm f} = \pi/\Omega_{0,1}$, which leads to $p_1(t_{\rm f}) = 1$ if and only if $\omega_{0,1}^{(\rm d)} = \tilde{\omega}_{0,1}$. The populations shown in \ref{fig:dw02_dw} are the averaged state $\ket{1}$ populations during the simulation
\begin{equation}
\bar{p}_i = \frac{1}{t_{\rm f}}\int_0^{t_{\rm f}} |\braket{i}{\psi}|^2\,\mathrm{d}t,
\end{equation}
which for $i = 1$ smoothly reduces from 1/2 to 0 when the probe gets detuned from the transition.

\ref{fig:dw02_dw}b) shows the corresponding simulation probing the 1 \--- 2 transition. The initial state of the system is set to $\ket{2}$ so that when $\omega_{1,2}^{\rm (d)}$ is resonant with $\tilde{\omega}_{1,2}$, the state $\ket{1}$ gets populated and allows the comparison of the numerical solution to the perturbation theory approximation $\epsilon_{1,2}$. As expected from Eq. \eqref{eq:energy_shifts}, $\tilde{\omega}_{0,1}$ and $\tilde{\omega}_{1,2}$  shift to the opposite directions by an equal amount because $\lambda = 1$ in the simulation.

In \ref{fig:dw02_dw}c) we have applied the dynamical phase corrections given in Eq. \eqref{eq:phi_correction} to both the 0 \--- 1 probe and the two-photon drive. Because in the simulation the drive amplitudes are constant, we take $\tilde{\phi}_{0,1}(t) = \epsilon_{0,1}t$, which effectively shifts the probe frequency by $\epsilon_{0,1}$, resulting in $\omega_{0,1}^{\rm (d)} = \omega_{0,1} + \delta_{0,1} + \epsilon_{0,1}$. When $\delta_{0,1} = 0$, the drive is supposed to be resonant with $\tilde{\omega}_{0,1}$, which is also seen in the simulation as a peak in the population $\bar{p}_1$, thus proving that the perturbative phase correction manages to account for the ac-Stark shift caused by the two-photon pulse. \ref{fig:dw02_dw}d) verifies the corresponding result also for the 1 \--- 2 transition.

Next we show that the correction also works as a function of the two-photon pulse amplitude $\Omega_{0,2}$ in a system with $\lambda = \sqrt{2}$. In \ref{fig:dw_A}a) a weak probe acting on the 0 \--- 1 transition excites the system to state $\ket{1}$ when it is resonant with the ac-Stark shifted transition frequency $\tilde{\omega}_{0,1}$. When $\Omega_{0,2} = 0$, there is no ac-Stark shift and $\tilde{\omega}_{0,1} = \omega_{0,1}$. We have also applied a small additional detuning to the two-photon drive, $\delta_{0,2} - \Delta/2 = \Delta/60$ because of the avoided crossing resulting from the resonant driving of the 0 \--- 2 transition at $\delta_{0,2} = \Delta/2$. The solid black line shows the estimate of the ac-Stark shift $\epsilon_{0,1}$ given by Eq. \eqref{eq:energy_shifts}. In \ref{fig:dw_A}b) the same simulation is repeated for the 1 \--- 2 transition presenting the correspondence of $\epsilon_{1,2}$ to the numerical simulation. \ref{fig:dw_A}c) demonstrates the self-induced ac-Stark shift on the 0 \--- 2 transition. If $\lambda \neq 1$ the shifts of the 0 \--- 1 and 1 \--- 2 transitions do not cancel, resulting in a shift also in the 0 \--- 2 transition. When the detuning $\delta_{0,2}$ is swept, the shift $\epsilon_{0,2}/2$ can be determined from the peak in the state $\ket{2}$ population. The observed shift is $\epsilon_{0,2}/2$ because the two-photon drive constitutes of two parts and the sum of their frequencies drives the transition. Here the simulation duration is $t_{\rm f} = \pi/|\Omega_{\rm eff}|$, so that the averaged state $\ket{2}$ population is $\bar{p}_2 = 0.5$ when the two-photon drive is resonant with $\tilde{\omega}_{0,2}/2$. The value of the shift is well captured by the approximation of Eq. \eqref{eq:energy_shifts}. 

%
%

\begin{figure}[tb]
	\centering
	\includegraphics[width=1.0\textwidth]{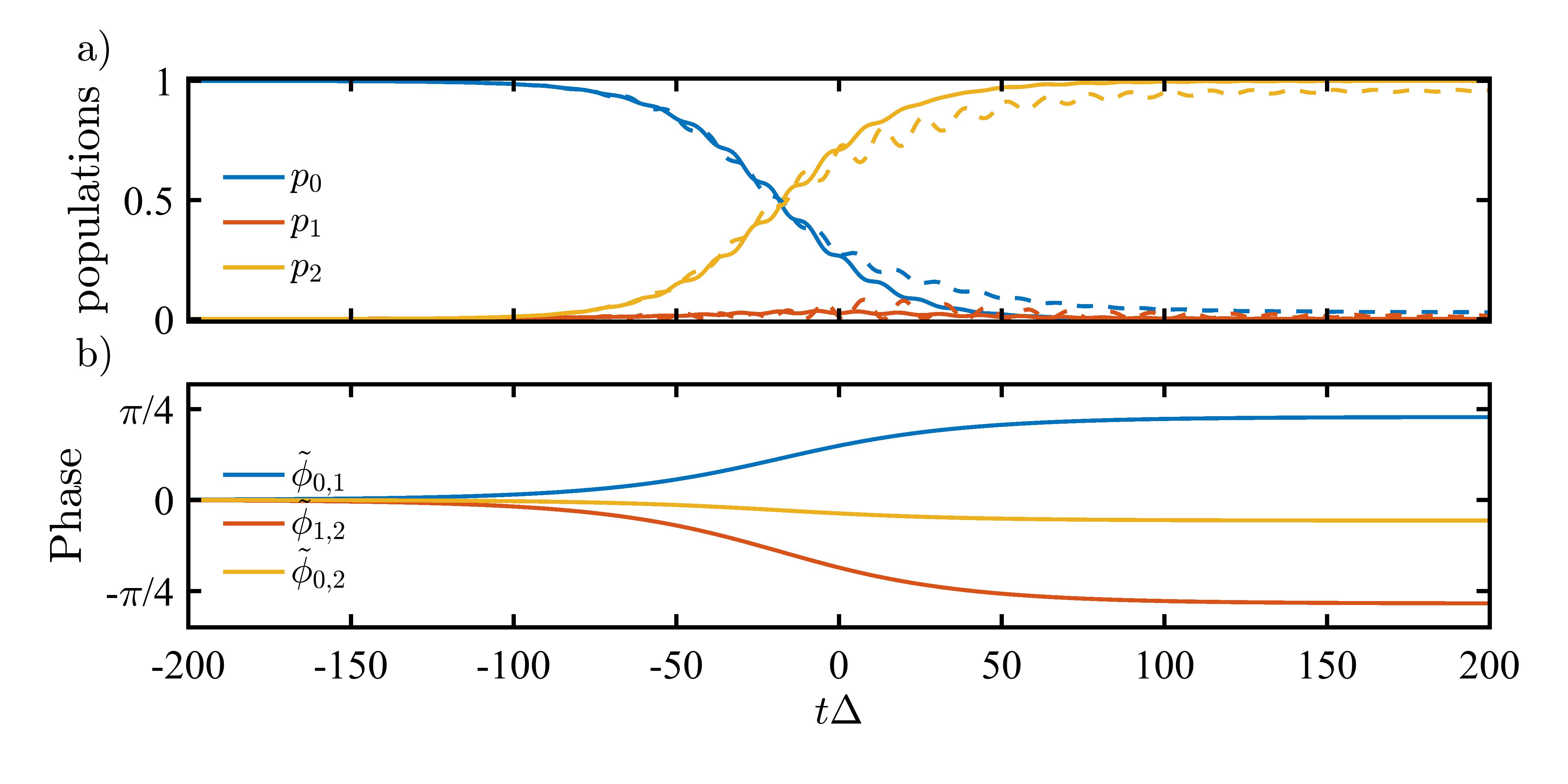}
	\caption{
		a) The time evolution of the occupation probabilities $p_0$, $p_1$ and $p_2$ during saSTIRAP with the system initially in the ground state. The dashed line corresponds to the case where $\phi_{0,2}(t) = \pi/4 + {\rm const}$, whereas the solid line is the simulation with the dynamic phase correction following Eqs. \eqref{eq:energy_shifts}, {\it i.e.} $\phi_{0,1}(t) = \tilde{\phi}_{0,1}(t)$, $\phi_{1,2}(t) = \tilde{\phi}_{1,2}(t)$, and $\phi_{0,2}(t) = \pi/4 + \tilde{\phi}_{0,2}(t)/2$. The small remaining oscillations in the populations result from the minor mixing of the states $\ket{0}$, $\ket{1}$ and $\ket{2}$ due to the two-photon driving. The dynamical phase correction ensures that the system evolves adiabatically in the dressed state basis. b) The dynamical phases $\tilde{\phi}_{0,1}(t)$, $\tilde{\phi}_{1,2}(t)$, and $\tilde{\phi}_{0,2}(t)$ used to compensate for the ac-Stark shifts. The parameters in the simulation are $\Omega_{0,1} = \Omega_{1,2} = \Delta/10$, $\lambda = \sqrt{2}$, $t_{\rm s} = -2\sigma$, and $\sigma = 36\Delta^{-1}$. With these numbers the area of the STIRAP pulses are $\mathcal{A} \approx 3\pi$. The peak counterdiabatic pulse amplitude is $\abs{\Omega_{0,2}} \approx \Delta/7$.
	}
	\label{fig:two_photon_correction}
\end{figure}


Above we have shown that the phases $\tilde{\phi}_{i,j}$ can cancel the ac-Stark shifts caused by the two-photon driving so that we can use the Hamiltonian $H_{\rm eff}$ from Eq. \eqref{eq:H_eff} to realize the saSTIRAP Hamiltonian $H + H_{\rm cd}$. By taking $\Omega_{\rm eff} = \Omega_{\rm cd}$ and $\Omega_{0,1} = \Omega_{1,2}$, we can use Eqs. \eqref{eq:cd} and \eqref{eq:effective_coupling} to calculate the two-photon driven counterdiabatic pulse shape
\begin{equation}
\label{eq:two_photon_pulse_shape}
|\Omega_{0,2}(t)| = \sqrt{-\frac{2 t_{\rm s}\delta_{0,2}}{\lambda\sigma^2\cosh\left[-\frac{t_{\rm s}}{\sigma^2}\left(t - t_{\rm s}/2\right)\right]}}
\end{equation}
and phase
\begin{equation}
\label{eq:two_photon_phase}
\phi_{0,2}(t) = \left[\arg{(\Omega_{\rm cd})} - \pi\right]/2 = -\pi/4,
\end{equation}
where the additional $\pi$ in the phase comes from the minus sign in the effective coupling.
\ref{fig:two_photon_correction}a) shows the time evolution of the populations $p_0$, $p_1$ and $p_2$ corresponding to the occupation probabilities of the states $\ket{0}$, $\ket{1}$ and $\ket{2}$ during saSTIRAP. The dashed lines show the populations when no dynamical corrections are applied to the drives. In this case ac-Stark shifts of all the three drives induce accumulated phases, which modify the optimal saSTIRAP phase $\phi_{0,2}$. This is taken into account by numerically finding the optimal (constant) phase $\phi_{0,2}$ that generates the highest fidelity, but even this leads to an imperfect result. The solid lines show that this is rectified by the dynamical phase correction following Eq. \eqref{eq:phi_correction}, which leads to $p_2 = 1$. The time evolution of the phase corrections is shown in \ref{fig:two_photon_correction}b). The result demonstrates that if the ac-Stark shift due to the two-photon driving is taken into account accordingly, it is possible to use the superadiabatic method with high fidelity in a three-level system with a forbidden transition. Here we note that because the phase correction is applied to the drives, the mixing of the states in the basis $\{\ket{0},\ket{1},\ket{2}\}$ leads to a small but not completely negligible population in state $\ket{1}$. However, due to the corresponding phase correction, this does not lead to reduction in fidelity. In the simulation the area of both STIRAP pulses was $\mathcal{A} = \int_{-\infty}^\infty \Omega_{0,1}(t) \mathrm{d}t = \int_{-\infty}^\infty \Omega_{1,2}(t) \mathrm{d}t \approx 3\pi$.

\subsection{Construction of NOT gates using superadiabatic STIRAP}
STIRAP in its standard form can transform the system from an initial state to a target state. Reversal of STIRAP has been demonstrated in \cite{stirap_ours}, as well as its operation under hybrid pulses \cite{hybrid_stirap}. In general, STIRAP cannot perform population inversion for an arbitrary state, because the adiabatic process depends on the initial state. However, it is known that by operating STIRAP with single photon detuning $\delta_{0,1} = -\delta_{1,2} = \delta \neq 0$ STIRAP works with either intuitive or counter-intuitive pulse sequence \cite{stirap_review_vitanov,du2014experimental}. Changing the pulse order is equivalent to switching the initial and the final states; this implies that off-resonant STIRAP can transfer the population either from state $\ket{0}$ to state $\ket{2}$ or vice versa. The same applies also to an arbitrary superposition in the $\{\ket{0},\ket{2}\}$ subspace, enabling population inversion of an unknown state. 
The operation of the reverse STIRAP can be understood from the instantaneous eigenstates of the system given in Eq. \eqref{eq:stirap_eig}. Similarly to resonant STIRAP, the initial state corresponding to $\ket{0}$ in the instantaneous basis is $\ket{\psi_{\rm i}} = \ket{D}$. The difference with respect to the resonant STIRAP emerges if the initial state in the original basis is $\ket{2}$, which leads to the instantaneous initial state being $\ket{\psi_{\rm i}} = \ket{-} = \ket{B}$ because $\Phi \approx 0$. In the resonant case the corresponding state would be $\ket{\psi_{\rm i}} = \frac{1}{\sqrt{2}}(\ket{+} + \ket{-})$. Even though $\Phi$ might become temporarily non-zero during the STIRAP pulses, as long as the process is adiabatic there are no diabatic excitations in the system, eventually resulting into the final state $\ket{\psi_{\rm f}} = \ket{-} = \ket{B} = \ket{0}$. Note that because $\Phi \approx 0$ only needs to be satisfied at the beginning and at the end of the pulse sequence when the amplitudes $\Omega_{0,1}(t)$ and $\Omega_{1,2}(t)$ are small, the condition on the detuning is much less severe than the requirement of the method introduced in \cite{Du2016}; there $\delta \gg \sqrt{\Omega_{0,1}^2(t) + \Omega_{1,2}^2(t)}$ needs to be satisfied always so that the contribution of state $\ket{1}$ can be adiabatically eliminated. The initial state of the system might also be a superposition state $\ket{\psi_{\rm i}} = \alpha\ket{0} + \gamma\ket{2}$, which in the instantaneous basis leads to a superposition $\ket{\psi_{\rm i}} = \alpha\ket{D} + \gamma\ket{B}$. In the adiabatic limit both of the states follow their own adiabatic paths, resulting in $\ket{\psi_{\rm f}} = \gamma\ket{0} - \alpha\ex{-i\phi_{\rm f}}\ket{2}$, where $\phi_{\rm f}$ is a phase factor accumulated because $\ket{-}$ is not a zero energy state unlike $\ket{D}$. Consequently, the off-resonant STIRAP realizes a unitary gate 
\begin{equation}
\label{eq:unitary}
U = \ket{0}\bra{2}\ex{i(\phi_{\rm f}+\pi)/2} + \ket{2}\bra{0}\ex{-i(\phi_{\rm f}+\pi)/2},
\end{equation}
up to an irrelevant common phase factor. If the phase $\phi_{\rm f}$ were 0, the operation could be considered a robust NOT gate, but as $\phi_{\rm f}$ strongly depends on the STIRAP pulse amplitudes the final state is sensitive to their fluctuations. Therefore off-resonant STIRAP can only guarantee robust population inversion. In \cite{fractional_gate_stirap} it has been suggested that the accumulated phase can be cancelled by employing two f-STIRAPs in series with the pulse order of the first f-STIRAP reversed. As a result, the phase accumulated during the second f-STIRAP has an opposite sign with respect to the first f-STIRAP leading to the cancellation of the accumulated phases.

The drawback of the off-resonant STIRAP or f-STIRAP is that the effective STIRAP amplitude is reduced because of the detuning $\delta$, thus impairing the operation fidelity. This cannot be completely avoided by increasing the peak STIRAP amplitudes $\Omega_{0,1}$ or $\Omega_{1,2}$, because larger amplitudes also require larger detuning in order to avoid the diabatic excitations at the beginning and at the end of the pulse sequence. As a result, longer STIRAP pulses are required to maintain the adiabaticity of the process. Consequently, the superadiabatic method is here particularly useful for enabling faster operation without decrease in fidelity.
In \cite{sa_stirap_gate} fractional STIRAP has been applied in a four-level tripod system to create a superadiabatic gate. In this work we demonstrate how two-photon driving enables implementation of a superadiabatic gate in a three-level ladder system.

In order to derive the counter-diabatic Hamiltonian in case $\delta \neq 0$, we substitute the eigenstates from Eq. \eqref{eq:stirap_eig} to the expression for the counterdiabatic Hamiltonian in Eq. \eqref{eq:re_ham}. After tedious but straightforward algebra one finds the new counter-diabatic Hamiltonian


\begin{figure}[tb]
	\centering
	\includegraphics[width=1.0\textwidth]{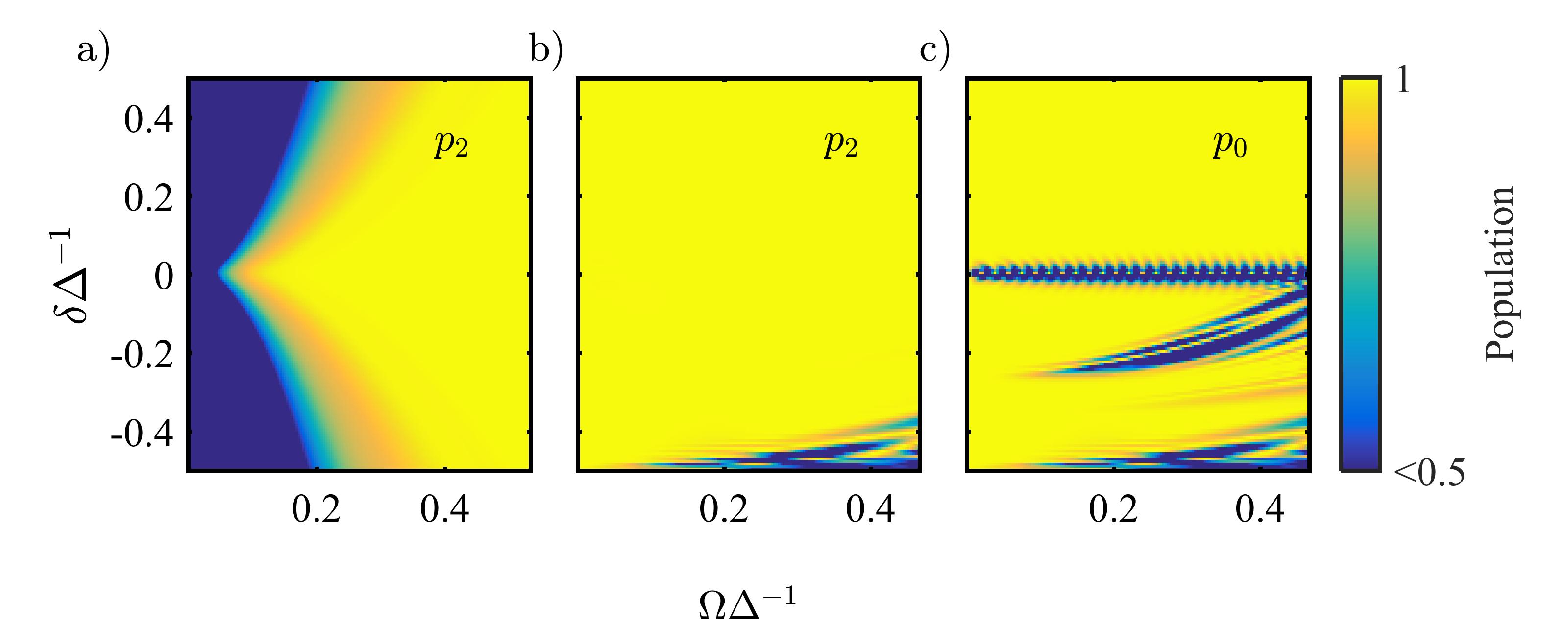}
	\caption{a) Population $p_2$ after STIRAP as a function of the single photon detuning $\delta$ and the STIRAP amplitude $\Omega = \Omega_{0,1} = \Omega_{1,2}$ when the initial state of the system is $\ket{0}$. b) Population $p_2$ after saSTIRAP starting from state $\ket{0}$. c) The population $p_0$ after saSTIRAP when the initial state is $\ket{2}$. STIRAP has parameters $\sigma = 80\Delta^{-1}$, $\lambda=1$, and $t_{\rm s} = -2\sigma$ in all the panels.}
	\label{fig:delta_vs_A01}
\end{figure}

\begin{equation}
\label{eq:cd_delta}
H_{\rm cd}^{(1)} = i\hbar\begin{bmatrix}
0 & \dot{\Phi}(t)\sin{\Theta(t)} & \dot{\Theta}(t) \\
-\dot{\Phi}(t)\sin{\Theta(t)} & 0 & -\dot{\Phi}(t)\cos{\Theta (t)} \\
-\dot{\Theta}(t) & \dot{\Phi}(t)\cos{\Theta(t)}  & 0 \\
\end{bmatrix},
\end{equation}
where, notably, the $0-2$ coupling maintains the same form as in Eq. \eqref{eq:cd}, but one needs to introduce small corrections to the $0-1$ and $1-2$ pulses with a phase difference of $\pi/2$ with respect to the phases of $\Omega_{0,1}(t)$ and $\Omega_{1,2}(t)$.
Without further modifications, saSTIRAP works well with single-photon detuning, which is demonstrated by the numerical simulation in \ref{fig:delta_vs_A01}b). There we show the population of state $\ket{2}$ as a function of the STIRAP amplitude $\Omega = \Omega_{0,1} = \Omega_{1,2}$ and the single photon detuning $\delta$. Even though the adiabaticity of STIRAP drops as $\delta$ is increased, the counterdiabatic pulse ensures that the population of state $\ket{2}$ remains 1. This can be compared to STIRAP without counterdiabatic correction in \ref{fig:delta_vs_A01}a) where $p_2$ drops as $\delta/\Omega$ grows.
Interestingly, unlike STIRAP, saSTIRAP is not symmetric with respect to $\delta$ because for large negative detunings both the $0-1$ and $1-2$ drive frequencies start approaching the two-photon drive frequency, causing interference. Fortunately this can be avoided by only considering the case where $\delta \ge 0$. In \ref{fig:delta_vs_A01}c) we show the population of state $\ket{0}$ with the initial state of the system being $\ket{2}$. As expected, with $\delta \approx 0$, the $1-2$ pulse resonantly transfers population to $\ket{1}$ invalidating the adiabatic process. However, as $\delta$ is increased, the adiabatic behaviour is regained.


\begin{figure}[tb]
	\includegraphics[width=1.0\textwidth]{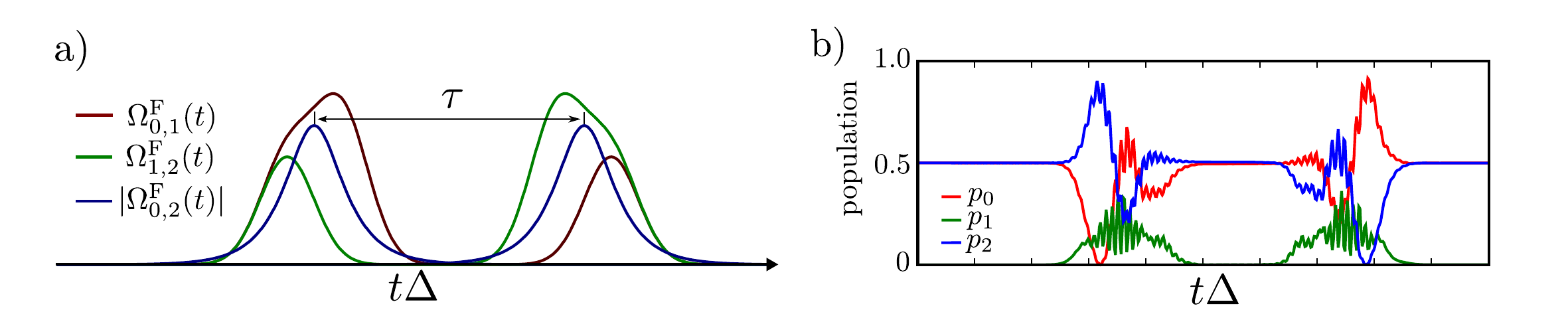}\\
	\includegraphics[width=1.0\textwidth]{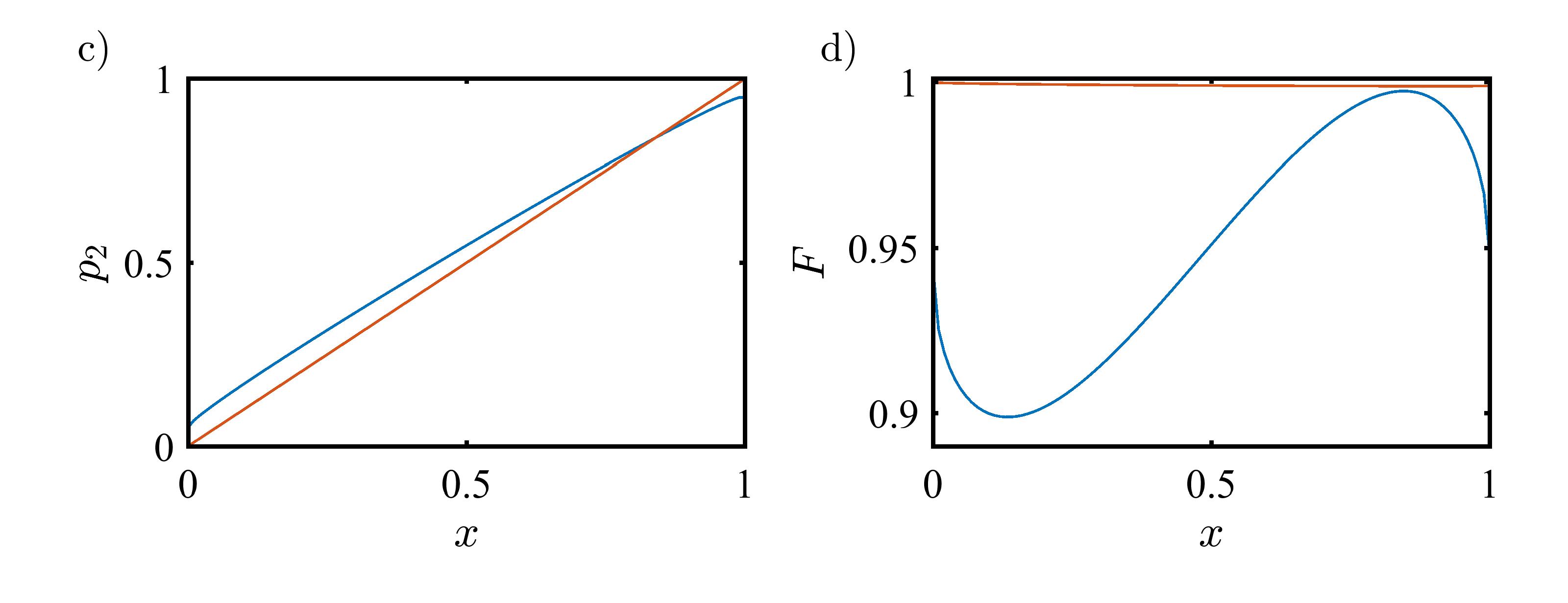}
	\caption{a) Pulse shapes used to create superadiabatic NOT gate. The pulse sequence consists of two fractional saSTIRAPs. b) Populations during the superadiabatic NOT gate for the initial state $\ket{\psi_{\rm i}} = (\ket{0} + i\ket{2})/\sqrt{2}$. c) State $\ket{2}$ population $p_2$ as a function of the initial state $\ket{\psi_{\rm i}(x)}$ after adiabatic NOT gate (blue) and superadiabatic NOT gate (red) with $\Omega_{0,1} = \Omega_{1,2} = \Omega = \Delta/6$, $\sigma = 36\,\Delta^{-1}, \delta = 0.1\Delta$, $\lambda = 1$, and $t_{\rm s} = -2 \sigma$. With these parameters the pulse area of the full STIRAP is $\mathcal{A} \approx 5 \pi$ and the peak amplitude of the counterdiabatic pulse is $\abs{\Omega_{0,2}^{\rm F}} = 0.24\Delta$. d) The corresponding fidelities calculated using Eq. \eqref{eq:fidelity}.
	}
	\label{fig:sa_stirap_gate}
\end{figure}
In order to employ the detuned saSTIRAP as a unitary gate, we divide the operation into two fractional saSTIRAPs (f-saSTIRAP), which cancel the accumulated dynamical phases. Following \cite{fractional_gate_stirap}, the combined f-STIRAP pulse shapes are

\begin{equation}
\begin{aligned}
\Omega_{0,1}^{\rm F}(t) &= \Omega_{0,1}(t) + \cos(\eta)\Omega_{1,2}(t) + \sin(\eta)\Omega_{0,1}(t - \tau), \\
\Omega_{1,2}^{\rm F}(t) &= \sin(\eta)\Omega_{1,2}(t) + \Omega_{1,2}(t - \tau) + \cos(\eta)\Omega_{0,1}(t - \tau),
\end{aligned}
\label{eq:fractional_stirap_pulse_shapes}
\end{equation}
where $\tau = 10\sigma$ is the delay between the two f-STIRAPS, $\Omega_{0,1}(t)$ and $\Omega_{1,2}(t)$ are given in Eqs. \eqref{eq:pulse_shapes}, and $\eta$ determines the fraction of the STIRAP. Here the initial mixing angle $\tan(\Theta_{\rm i}) = \Omega_{0,1}^{\rm F}(t_{\rm i})/\Omega_{1,2}^{\rm F}(t_{\rm i}) = \cos(\eta)/\sin(\eta) = \cot(\eta)$ mixes the states $\ket{D}$ and $\ket{-} \approx \ket{\rm B}$ so that $\ket{\psi_{\rm i}} = \alpha\ket{0} + \gamma\ket{2} = \braket{\rm D(\Theta_{\rm i})}{\psi_{\rm i}}\ket{\rm D(\Theta_{\rm i})} + \braket{\rm B(\Theta_{\rm i})}{\psi_{\rm i}}\ket{\rm B(\Theta_{\rm i})}$ in the adiabatic basis. After the first f-STIRAP the intermediate mixing angle is $\Theta_{\rm int}^{(1)} = \pi/2$ and the state adiabatically evolves to $\ket{\psi_{\rm int}} = \left[\alpha\cos(\eta) + \gamma\sin(\eta)\right]\ex{-i\phi_{\rm f}}\ket{0} + \left[\gamma\cos(\eta) - \alpha\sin(\eta)\right]\ket{2}$, where $\phi_{\rm f}$ is the accumulated phase and we have used $\ket{\rm D(\Theta_{\rm int}^{(1)})} = -\ket{2}$ and $\ket{\rm B(\Theta_{\rm int}^{(1)})} = \ket{0}$. The initial mixing angle for the second f-STIRAP is $\Theta_{\rm int}^{(2)} = 0$, which evolves to $\tan(\Theta_{\rm f}) = \sin(\eta)/\cos(\eta) = \tan(\eta)$, resulting in state $\ket{\psi_{\rm f}} = i\left[\alpha\cos(2\eta) + \gamma\sin(2\eta)\right]\ex{-i(\phi_{\rm f} + \pi/2)}\ket{0} + i\left[\gamma\cos(2\eta) - \alpha \sin(2\eta)\right]\ex{-i(\phi_{\rm f} + \pi/2)}\ket{2}$, where the common phase factor $\ex{-i(\phi_{\rm f} + \pi/2)}$ can be ignored. For $\eta = \pi/4$ this results in $\ket{\psi_{\rm f}} = i\gamma\ket{0} - i\alpha\ket{2}$ and therefore the unitary that realizes the transformation $\ket{\psi_{\rm i}} \rightarrow \ket{\psi_{\rm f}}$ is given by
\begin{equation}
\label{eq:not_unitary}
U = \ex{i\pi/2}\ket{0}\bra{2} + \ex{-i\pi/2}\ket{2}\bra{0},
\end{equation}
which is a NOT-gate independent of the accumulated phase $\phi_{\rm f}$.

The counterdiabatic pulse shape $\Omega_{\rm cd}^{\rm F}(t)$ required to realize the superadiabatic fractional STIRAP is derived by substituting Eqs. \eqref{eq:fractional_stirap_pulse_shapes} into Eq. \eqref{eq:cd_delta}. Furthermore, the actual two-photon pulse amplitude $\Omega_{0,2}^{\rm F}(t)$ and the phase $\phi_{0,2}^{\rm F}$ corresponding to the counterdiabatic pulse can be calculated from Eq. \eqref{eq:effective_coupling} similarly to standard saSTIRAP. The resulting pulse shapes are shown in \ref{fig:sa_stirap_gate}a). \ref{fig:sa_stirap_gate}b) shows the time evolution of the populations during a NOT gate when the initial state of the system is $\ket{\psi} = (\ket{0} + i\ket{2})/\sqrt{2}$ and the parameters of the simulation are $\Omega = \Omega_{0,1} = \Omega_{1,2} = \Delta/6$, $\delta = 0.1\Delta$ and $\sigma = 36\Delta^{-1}$.
The gate reliably realizes population inversion for any initial state 
\begin{equation}
\label{eq:initial_state}
\ket{\psi_{\rm i}(x)} = \sqrt{x}\ket{0} + i\sqrt{1-x}\ket{2}
\end{equation}
with $x \in [0,1]$, which is shown in \ref{fig:sa_stirap_gate}c). The figure demonstrates the importance of the superadiabatic correction by showing the populations  $p_2$ after a f-saSTIRAP gate (red line) or f-STIRAP gate (blue line). Without the counterdiabatic correction, the populations significantly deviate from the straight line. This is further demonstrated in \ref{fig:sa_stirap_gate}d), where we show the gate fidelity for the two cases. The fidelity is defined as \cite{NielsenChuang}
\begin{equation}
\label{eq:fidelity}
F(\psi_{\rm i}) = \bra{\psi_{\rm i}}U^\dagger A(\psi_{\rm i}) U\ket{\psi_{\rm i}},
\end{equation}
where $U$ is given in Eq. \eqref{eq:not_unitary} and $A = \ket{\psi_{\rm f}}\bra{\psi_{\rm f}}$, which becomes simply $F(\psi_{\rm i},\psi_{\rm f}) = |\bra{\psi_{\rm f}}U\ket{\psi_{\rm i}}|^2$. For the f-saSTIRAP gate, the fidelity is almost 1 for all $x$, whereas f-STIRAP fidelity is significantly lower.

The simulation proves that it is possible to realize an adiabatic NOT gate for an arbitrary initial state using saSTIRAP, but next we will show that the operation also retains the robustness usually associated with adiabatic methods, which is supposed to be far superior when compared to a gate realized with a standard $\pi$-pulse.

\begin{figure}[tb]
	\centering
	\includegraphics[width=1.0\textwidth]{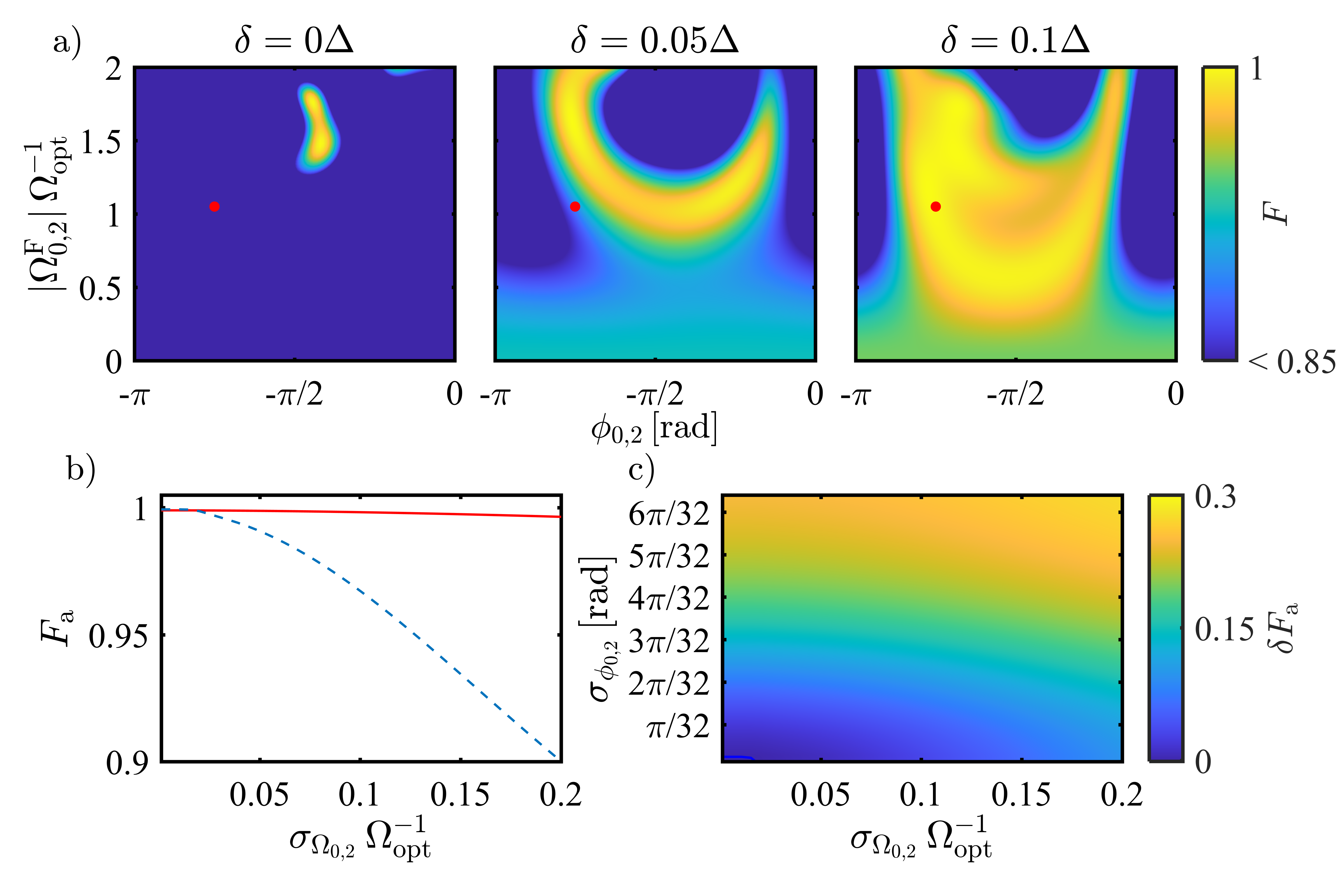}
	\caption{a) The gate fidelity $F$ as a function of the two-photon drive amplitude $|\Omega_{0,2}^{\rm F}|$ and phase $\phi_{0,2}^{\rm F}$ for the initial state $x = 1$. The amplitude values are normalized with the optimal counterdiabatic drive amplitude $\Omega_{\rm opt} = 0.24\Delta$. The panels show the gate fidelity with increasing values of the detuning $\delta$. The optimal working point for saSTIRAP is shown with the red dot.
		b) The averaged gate fidelity of saSTIRAP (solid red line) is compared to the direct two-photon $\pi$-pulse (dashed blue line) having the same shape as the counter-diabatic pulse. The amplitudes of the $\pi$-pulse and the counterdiabatic pulse fluctuate around their optimal values $\Omega_{\rm opt} = 0.24\Delta$ with the standard deviation $\sigma_{\Omega_{0,2}}$.
		The simulation is performed with $\delta = 0.1\Delta$. c) The panel shows the difference in fidelity $\delta F_{\rm a}$ between superadiabatic NOT gate and the direct two-photon $\pi$-pulse in the presence of fluctuations in pulse amplitude $|\Omega_{0,2}^{\rm F}|$ with standard deviation $\sigma_{\Omega_{0,2}}$ and phase $\phi_{0,2}^{\rm F}$ with standard deviation $\sigma_{\phi_{0,2}}$. Here the saSTIRAP fidelity is higher for all the values of fluctuations except for the zero fluctuation case where the fidelities are equal.
		The STIRAP parameters used in all the panels are $\Omega_{0,1} = \Omega_{1,2} = \Omega = \Delta/6$, $\sigma = 36\,\Delta^{-1}$, $\lambda=1$, and $t_{\rm s} = -2 \sigma$. The pulse area of a full STIRAP is $\mathcal{A} \approx 5\pi$. 
	}
	\label{fig:robustness}
\end{figure}

We start by numerically calculating how the gate performs as a function of the counter-diabatic pulse parameters. In \ref{fig:robustness}a) we show the gate fidelity when the maximum of the two-photon amplitude $|\Omega_{0,2}^{\rm F}|$ and phase $\phi_{0,2}^{\rm F}$ deviate from their optimal values $\Omega_{\rm opt}$ or $-\pi/4$ for several values of the single photon detuning.
If $\delta$ is too small, the fidelity of the gate remains low because f-saSTIRAP with reversed pulse order does not work as intended.
To quantitatively analyze the robustness of the gate we fix $\delta = 0.1\Delta$ and  simulate the fluctuations of the control parameters with normally distributed random variables $|\Omega_{0,2}^{\rm F}| \sim  \mathcal{N}(\Omega_{\rm opt},\sigma_{\Omega_{0,2}})$ and $\phi_{0,2}^{\rm F} \sim \mathcal{N}(-\pi/4,\sigma_{\phi_{0,2}})$ and calculate the averaged fidelity weighted over their distributions

\begin{equation}
\label{eq:avg_fidelity}
F_{\rm a} = \iiint \df\psi_{\rm i}\df v_1\df v_2 F(\psi_{\rm i},\psi_{\rm f}(v_1,v_2)) P(|\Omega_{0,2}^{\rm F}| = v_1) P(\phi_{0,2}^{\rm F} = v_2).
\end{equation}
Here $P(V = v)$ is the probability to get a value $v$ for the random variable $V$ and $\ket{\psi_{\rm f}(v_1,v_2)}$ is the state of the system after the gate, which depends on the values of $|\Omega_{0,2}^{\rm F}|$ and $\phi_{0,2}^{\rm F}$ as shown in \ref{fig:robustness}a) for the initial state $\ket{\psi_{\rm i}} = \ket{0}$.
The integration should be performed over all the possible initial states $\ket{\psi_{\rm i}}$ in order to guarantee that the averaged fidelity properly includes an arbitrary initial state. Due to the numerical complexity we here include only the states given by $\ket{\psi_{\rm i}(x)}$ in Eq. \eqref{eq:initial_state}.


\ref{fig:robustness}b) shows the effect of fluctuations in 0\---2 drive amplitude on the gate fidelity calculated with different values of the fluctuation strength $\sigma_{\Omega_{0,2}}$ in the absence of phase fluctuations.
When compared to the $\pi$-pulse realized with two-photon driving of the 0\---2 transition, the superadiabatic NOT gate provides a dramatic improvement in fidelity because its robustness protects it against the fluctuations in amplitude. In \ref{fig:robustness}c) we also take into account the fluctuations of the phase of the counterdiabatic pulse and compare the resulting averaged fidelity to the case where the 0\---2 $\pi$-pulse is subjected to the same fluctuations. In this case the superiority of the adiabatic method becomes even more prominent, demonstrating that the addition of the counterdiabatic pulse does not render the superadiabatic method susceptible to noise in the control parameters.
However, it is important to note that the actual rotation angle of the adiabatic gate is determined by the sum of the phases of the STIRAP pulses (which in this work are assumed to be zero), whose fluctuations are not included in the simulation. Instead, here we have shown that the counterdiabatic correction is not sensitive to either amplitude or the phase fluctuations.

In the analysis we have not included the fluctuations in the STIRAP amplitudes, even though they can potentially have influence on the fidelity of the process. However, we argue that their influence is small because the counterdiabatic correction is independent of the STIRAP amplitudes as long as they are equal, as shown in Eq. \eqref{eq:cd}. Experimentally the assumption of correlated fluctuations is feasible, because using amplitude modulation the pulses could be created with a single channel of an arbitrary waveform generator, which would inflict the same noise on both of the amplitudes.





\section{Conclusions}
In this article we have shown that two-photon driving can be used to realize superadiabatic STIRAP even in systems with forbidden transitions. However, in order to compensate for the ac-Stark shifts induced by the two-photon drive, a dynamical phase correction to all the drive pulses is required. With the corrections applied, we have shown that saSTIRAP can prepare state $\ket{2}$ with unit fidelity.

Next we have shown that it is possible to employ saSTIRAP to perform robust population inversion in a system initially in an arbitrary or unknown state. This is accomplished by applying single-photon detuning to STIRAP pulses and applying a corresponding counterdiabatic pulse. In contrast to STIRAP, where single photon detuning leads to a loss in fidelity unless the pulse duration is increased, the population inversion realized with saSTIRAP has unit fidelity. 
This can be further applied to implement a NOT gate in $\{\ket{0},\ket{2}\}$ subspace by combining two fractional saSTIRAPs, which cancel the accumulated dynamical phase during the gate. With the two-photon driving and the dynamical phase-correction, the population inversion as well as the NOT gate can be readily realized in a transmon. Moreover, we have studied the robustness of the superadiabatic NOT gate and compared it to a Rabi $\pi$-pulse, demonstrating that superadiabatic method provides significant protection from the fluctuations in the control parameters.

\section*{References}

\bibliographystyle{iopart-num}
\bibliography{ref}


\noindent{\bf Acknowledgements}\\
We acknowledge financial support from Väisälä Foundation, the Academy of Finland (project 263457), the Center of Excellence ''Low Temperature Quantum Phenomena and Devices'' (project 250280) and Aalto Centre for Quantum Engineering (projects QMET and QMETRO).

\noindent{\bf Competing financial interests:} The authors declare no competing financial interests.

\end{document}